\documentclass[%
reprint,
superscriptaddress,
amsmath,amssymb,
aps,
prb,
floatfix
]{revtex4-2}

\usepackage{multirow}
\usepackage{graphicx}
\usepackage{dcolumn}
\usepackage{bm}
\usepackage[version=3]{mhchem}
\usepackage{nicefrac}

\usepackage{color}
\usepackage{dcolumn}
\usepackage{amssymb}
\usepackage{url}
\usepackage{hyperref}
\usepackage{xfrac}
\usepackage{tabularx}
\usepackage{xspace}
\usepackage{amsmath}
\usepackage[normalem]{ulem}
\usepackage{mathtools}
\usepackage{physics}
\usepackage{dsfont}
\usepackage{float}

\def\be{\begin{equation}}
\def\ee{\end{equation}}
\def\bea{\begin{eqnarray}}
\def\eea{\end{eqnarray}}
\def\ba{\begin{array}}
\def\ea{\end{array}}

\begin{document}
\title{Chiral crossroads in \ce{Ho3ScO6}: a tale of frustration in maple leaf lattice}
\author{Pratyay Ghosh}
\email{pratyay.ghosh@epfl.ch}
\affiliation{Institute of Physics, Ecole Polytechnique Fédérale de Lausanne (EPFL), CH-1015 Lausanne, Switzerland}


\begin{abstract}
Motivated by the recent observation of a uniform vector chirality (UVC) magnetic order in the maple-leaf lattice (MLL) realization \ce{Ho3ScO6} via powder neutron scattering experiments, we investigate the classical antiferromagnetic Heisenberg model on the maple-leaf lattice. The MLL features three symmetry-inequivalent nearest-neighbor couplings, $J_d$, $J_t$, and $J_h$. Previous studies, primarily focused on the case where $J_t = J_h$, identified a staggered vector chirality (SVC) order. Extending beyond this limit, we demonstrate that the SVC order remains stable across a broad parameter regime. However, we also find that the UVC order cannot emerge from the nearest-neighbor model alone. By introducing a further-neighbor antiferromagnetic interaction, $J_x$, we demonstrate that even a weak $J_x$ can cause a first-order phase transition from SVC to UVC order. Using linear spin wave theory, we compute the dynamical spin structure factor, revealing distinct signatures for SVC and UVC orders that can be probed through inelastic neutron scattering experiments. Additionally, we calculate the specific heat, which exhibits qualitative agreement with the experimental data for \ce{Ho3ScO6}. Our findings provide a minimal framework for understanding \ce{Ho3ScO6} and related MLL systems, like \ce{MgMn3O7.3H2O}, suggesting avenues for further experimental and theoretical investigations.
\end{abstract}

\maketitle

\section{Introduction}
Frustrated magnetism is a field of study with both theoretical and experimental interests, centered around geometric constraints that hinder the system from simultaneously satisfying all pairwise interactions. When combined with quantum fluctuations, these constraints promote exotic magnetically disordered states, such as valence bond solids, spin liquids, spin nematics---complex and highly entangled spin states at low temperatures\cite{Diepbook,frustrationbook}. Even in the absence of quantum fluctuations—i.e., in classical spin systems—frustration alone can lead to magnetically ordered spin states which diverge significantly from conventional N\'eel-type antiferromagnetic magnetic ordering. Instead, a wide variety of non-collinear and non-coplanar long-range magnetic ordered ground states is observed, including incommensurate helical spin structures \cite{Rastelli1979,Sindzingre2009,Iqbal2016,Hasik2024}, conical spirals~\cite{Ghimire2020,PhysRevB.96.140401}, umbrella states~\cite{PhysRevB.93.224402,PhysRevB.91.024410}, canted antiferromagnetism ~\cite{Bukowski2022,Kipp2021}, and chiral magnetic orders~\cite{Xiong2020,PhysRevB.106.024401}. 

The triangular lattice and its derivatives serve as prototypical frustrated playgrounds for such phenomena to occur. For instance, the classical nearest-neighbor antiferromagnetic Heisenberg model on the kagome lattice, obtained by a $1/4$ site-depletion of the triangular lattice, exhibits a highly degenerate manifold of coplanar ground states with weathervane modes \cite{PhysRevLett.68.855,Schnabel2012-ky,PhysRevB.78.094423}. This degeneracy can be lifted through additional couplings—such as next-nearest-neighbor interactions or Dzyaloshinskii–Moriya interactions—leading to long-range magnetic order which can exhibit a vector chirality pattern across the lattice \cite{PhysRevLett.69.832, Ghosh-Spin-1_Kagome,PhysRevResearch.4.043019,PhysRevB.95.094427,PhysRevB.97.104419}.

\begin{figure*}[t]
\includegraphics[width=0.8\textwidth]{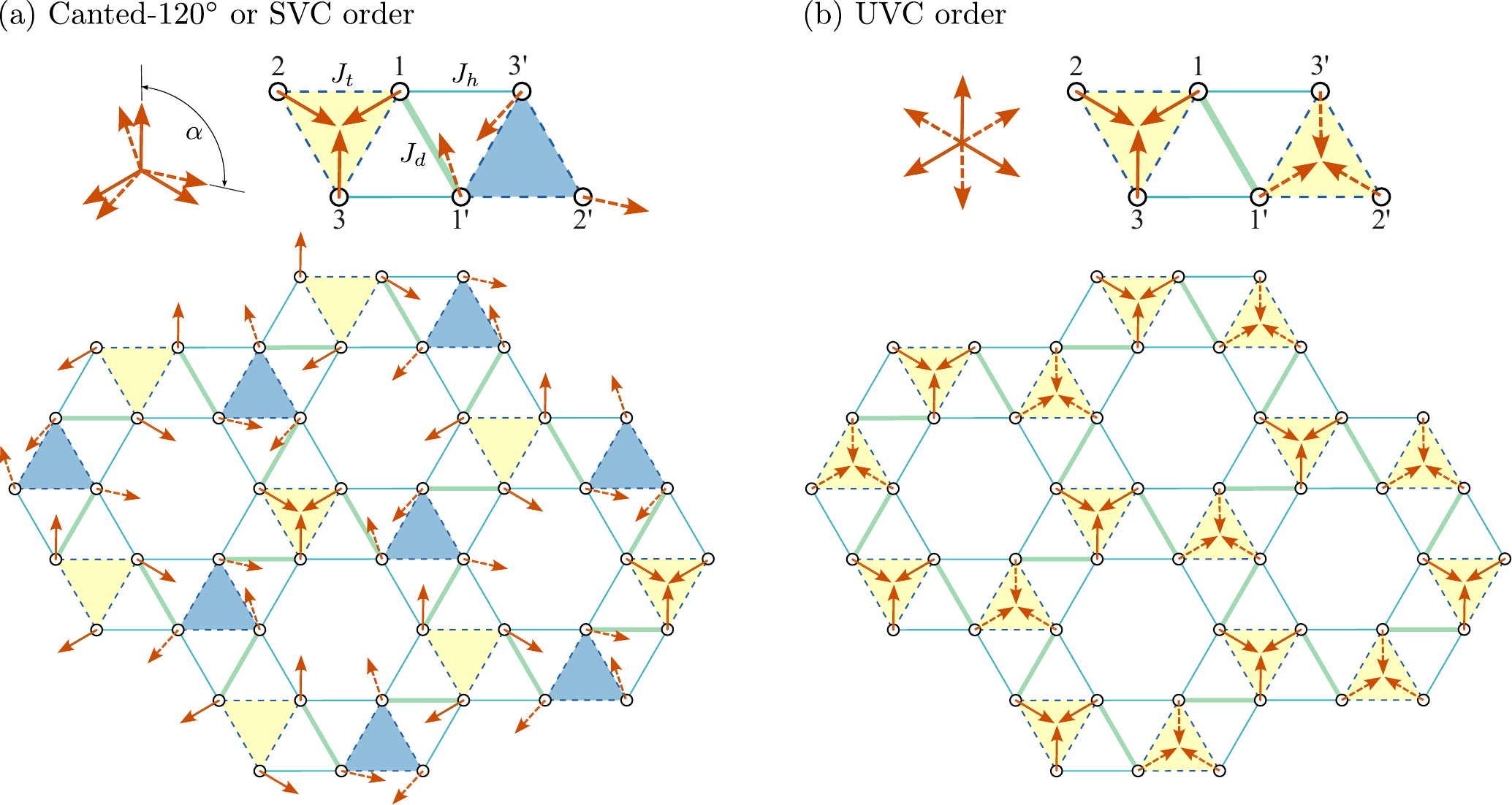}
\caption{The Maple-leaf lattice (MLL) with three symmetry-inequivalent bonds with exchange coupling $J_d$, $J_t$ and $J_h$. (a) The classical antiferromagnetic Heisenberg model on the MLL admits a canted-$120^\circ$ magnetic order for $J_t\ge J_h$~\cite{Ghosh2022}. In this order, the individual triangles accrue a $120^\circ$ spin confuration and there exists a relative canting, $\alpha$, between the spins on neighboring triangles leading to an enlargement of the unit-cell. The canting angle, $\alpha$, dpends upon the relative strength of $J_d$ and $J_h$ (see~\eqref{eq:canting_1}). This state exhibits a staggered arrangement of positive and negative vector chirality on the $J_t$-triangles. We also refer to the state as a staggered vector chirality (SVC) order. (b) In contrast, the \ce{Ho3ScO6}, a realization of a maple-leaf magnet, displays a state with uniform vector chirality (UVC) on all the $J_t$-triangles. Similar to the c-$120^\circ$ state, the UVC state also forms a $120^\circ$ spin arrangement on $J_t$-triangles, with spins across the $J_d$ bonds aligned antiparallel.} \label{fig-lattice}
\end{figure*}

Another candidate in this list is the maple-leaf lattice (MLL)~\cite{Betts1995} which has recently gained attention~\cite{Ghosh2022,PhysRevB.65.224405,Farnell2011,Farnell2014,Gresista2023,PhysRevB.110.014414, 
Ghosh2023,ghosh2024spinliquidmapleleafquantum,Ghosh2024-al,schmoll2024bathingseacandidatequantum} due to its kinship to the well-known Shastry-Sutherland lattice~\cite{Shastry1981} and several material realizations. Similar to the kagome lattice, the MLL is another example of a lattice derived from the triangular lattice through site depletion, specifically a 1/7-th site depletion. The MLL (Fig.~\ref{fig-lattice}) has a coordination number $5$ and belongs to $p6$ wallpaper group~\cite{PhysRevB.110.014414,Beck2024}. 

The MLL features three symmetry inequivalent nearest-neighbor bonds, distinguished by different colors and line types in Fig.~\ref{fig-lattice}. These bonds can be assigned with Heisenberg interactions, each with its own coupling constant: $J_h$ for the bonds forming the hexagons, $J_t$ for the bonds on the triangles, and $J_d$ for the dimer bonds. In this study, we focus on the fully antiferromagnetic model, i.e. $J_d,J_h,J_t>0$. For spin-$1/2$ system, and when $J_d/2 > J_t=J_h$, this model permits an analytical solution and admits an exact product dimer singlet groundstate. This makes the MLL the only other Archimedean lattice (as well as, lattices composed of uniform tiles) capable of doing so, alongside the Shastry-Sutherland model~\cite{Ghosh2022}. Furthermore, the classical ($S\to \infty$) Heisenberg antiferromagnet with $J_d/2 < J_t=J_h$, an exact evaluation of the ground state is possible via Luttinger-Tisza analysis~\cite{Luttinger1946,Kaplan2007}. The classical ground state comprises a local $120^\circ$ spin configuration on each $J_t$-triangle, with a non-local spin canting between the triangles that depends on both $J_d$ and $J_h$~\cite{PhysRevB.65.224405,Farnell2011,Ghosh2022}. Fig.~\ref{fig-lattice} (a) shows an example of such a magnetic order. This magnetic order is deemed as a canted-$120^\circ$ order in Ref.~\cite{Ghosh2022}. This coplanar $120^\circ$ structures also exhibits staggered vector chirality $\kappa\sim \left(\vec{S}_1\times\vec{S}_2+\vec{S}_2\times\vec{S}_3+\vec{S}_3\times\vec{S}_1\right)$ on each $J_t$-triangle. Due to this property, this order is also labeled as a staggered vector chirality order~\cite{Lake2025,Haraguchi2018}. Throughout this article, we will interchangeably refer to this state as c-$120^\circ$ order (short for canted-$120^\circ$ order) or staggered vector chirality (SVC) order.

The MLL has been reported to be realized in several magnetic systems, including bluebellite \ce{Cu6IO3(OH)10Cl}~\cite{Mills2014}, spangolite \ce{Cu6Al(SO4)(OH)12Cl.3H2O}~\cite{Hawthorne1993}, sabelliite \ce{Cu2ZnAsO4(OH)3}~\cite{Olmi1995},  mojaveite \ce{Cu6TeO4(OH)9Cl}~\cite{Mills2014}, fuetterite \ce{Pb3Cu6TeO6(OH)7Cl5}~\cite{Kampf2013}, \ce{MgMn3O7.3H2O}~\cite{Haraguchi2018}, and, most recently in \ce{Ho3ScO6}~\cite{Lake2025}. Among this, the magnetic properties for bluebellite \ce{Cu6IO3(OH)10Cl}~\cite{Ghosh2023b}, spangolite \ce{Cu6Al(SO4)(OH)12Cl.3H2O}~\cite{Fennell2011,schmoll2024tensornetworkanalysismapleleaf}, \ce{MgMn3O7.3H2O}~\cite{Haraguchi2018}, and \ce{Ho3ScO6}~\cite{Lake2025} have been investigated and understood to some extent. Both, \ce{MgMn3O7.3H2O} and \ce{Ho3ScO6} exhibit magnetic ordering at low-temperatures~\cite{Haraguchi2018,Lake2025}. Although this article primarily focuses on the properties of \ce{Ho3ScO6}, the discussions presented here may also be partially applicable to \ce{MgMn3O7.3H2O}. 

In their recent article, Lake \textit{et al.}~\cite{Lake2025} reports that the compound \ce{Ho3ScO6} consists of stacked MLL layers that develop a long-range magnetic order below N\'eel temperature $T_N = 4.1$ K with rather large \ce{Ho^{3+}} magnetic moments ($\sim 7 \mu_\text{B}$). The powder neutron diffraction data is consistent with a magnetic order comprised of $120^\circ$ spin arrangements on the $J_t$-triangles and a uniform vector chirality throughout the $ J_t$-triangles of the lattice. This spin configuration suggests a coupling hierarchy of $J_d\gg J_t>J_h$. In this scenario, the $J_d$ bonds are fully satisfied, and $J_h$ bonds remain significantly unsatisfied [Fig.~\ref{fig-lattice} (b)]. However, as we proceed in this article, it will become clear that any finite combination of $J_d\gg J_t>J_h$ would result in an SVC order, which starkly contrasts with the experimental findings. Therefore, it becomes apparent that other spin interactions must exist in \ce{Ho3ScO6} that compel it to realize a uniform vector chirality (UVC) order. We argue that possible weaker interactions, namely, additional further neighbor Heisenberg interactions, may be present in the system and could dictate the emergence of this particular order. Given the large magnetic moments of \ce{Ho^{3+}}, we restrict ourselves only to classical spins. We first examine the properties of the SVC and UVC orders. We then propose a further neighbor Heisenberg term that stabilizes the UVC order. In addition, we calculate the dynamical spin structure for our proposed model which can be compared with possible inelastic neutron scattering experiments to validate the model.

\section{Classical magnetic orderes in MLL antiferromagnet}
In this section, we will discuss the basic features of the coplanar classical c-$120^\circ$ order, also known as the staggered vector chirality (SVC) order, and the uniform vector chirality (UVC) order, assuming $J_d>J_t>J_h$. Apart from the usual degeneracy due to spin rotation, both orders have a two-fold degeneracy associated with the chirality, that is, the charalty of all the $J_t$-triangles can be switched simultaneously. The following discussions are independent of this degeneracy. 

\subsection{Canted-$120^\circ$ or Staggered Vector Chirality Order}
For the MLL antiferromagnet with $J_h=J_t$ model, studies in Refs.~\cite{Farnell2011,Ghosh2022} find the c-$120^\circ$ order for all values of $J_d$. In this order, each triangle accrues a $120^\circ$ spin configuration with either positive or negative vector charity, arranged in a staggered structure where each positive vector charity triangle is surrounded by three negative vector charity triangles and vice versa [see Fig.~\ref{fig-lattice} (a)]. In this order, the individual triangles are optimally satisfied, and there exists a relative canting between the spins on neighboring triangles, leading to an enlargement of the unit cell. 

It can be shown that this order is also realized for any combination of $J_h$ and $J_t$, as long as $J_h$ remains weaker than $J_t$.  The rationale behind this is that the $J_d$ interactions, on their own, do not introduce frustration into the system. As a result, it is possible to optimally satisfy the $J_t$-triangles and fully satisfy the $J_d$ bonds simultaneously. The addition of the $J_h$ bonds introduces frustration; however, due to the geometric constraints, it does not alter the $120^\circ$ spin configuration on individual $J_t$-triangles, but instead, induces the relative spin canting between the spins on neighboring triangles. The canting depends solely on $J_d$ and $J_h$, and can be parameterized by the relative angle formed between spins across the $J_d$ bonds as [see Fig.~\ref{fig-lattice} (a)]:
\be\label{eq:canting_1}
\alpha = \pi + \arctan \left[\frac{\sqrt{3} J_h}{J_h-J_d}\right].
\ee
The energy per site for this magnetic order is given by
\be
e_{\text{SVC}} = -\frac{J_t}{2} + J_h \cos(\frac{2 \pi}{3} + \alpha) + \frac{J_d}{2} \cos\alpha.
\ee
This energy formula highlights the complex interplay between $J_d$ and $J_h$ in setting up this order.

Note that, for the general case with $J_d>J_t>J_t$, the classical spin model does not allow for an exact solution via Luttinger-Tisza method~\cite{Ghosh2022}. Therefore, we resort to iterative energy minimization techniques to determine the ground state configurations. We employ a zero-temperature classical Monte Carlo energy minimization technique, wherein the spins are sequentially updated to locally minimize the system's energy. At each step, only energy-lowering updates are accepted, allowing the system to iteratively relax toward a ground state.
 
\subsection{Uniform Vector Chirality Order}
The uniform vector chirality (UVC) spin configuration is depicted in Fig.~\ref{fig-lattice}(b). We illustrate one specific UVC order where every $J_t$-triangle assumes a positive vector chirality. Note that while the spins on the hexagons also have a chirality, we do not refer to this while discussing the chirality of a certain order; the chirality mentioned here exclusively addresses the chirality associated with the $J_t$-triangles. Similar to the c-$120^\circ$ order, the UVC order also optimally satisfies the $J_t$-triangles, achieving a $120^\circ$ spin arrangement on each of them. However, unlike the c-$120^\circ$ state, the UVC order does not involve a spin canting that depends on the interaction strengths. Instead, the spins across the $J_d$ bonds remain strictly antiparallel. 

The lack of spin canting in the UVC state simplifies its energy evaluation compared to the SVC state. The energy per site for this state can be expressed as
\be
e_{\text{UVC}} = -\frac{J_t}{2} + J_h \cos\frac{\pi}{3} - \frac{J_d}{2}.
\ee
This energy formula indicates that the UVC order is stabilized by the $J_d$ and $J_t$ interactions, and distabilized by the $J_h$ interactions.

A comparison of energies, $e_{\text{SVC}}$ and $e_{\text{UVC}}$, reveals that the SVC consistently has a lower energy than the UVC order, except in the special case where $J_h=0$. At this limit, the energies of the SVC and UVC states become degenerate. However, this degeneracy extends far beyond just these two states: at $J_h = 0$, the classical ground state of the maple-leaf lattice Heisenberg antiferromagnet becomes extensively degenerate, closely related to the classical kagome Heisenberg antiferromagnet; the SVC and UVC orders are merely two members of the degenerate manifold.

\section{Spin Hamiltonian for \NoCaseChange{\ce{Ho3ScO6}}}
In this section, we aim to develop a minimal spin Hamiltonian that effectively describes the appearance of the UVC order observed in \ce{Ho3ScO6}. To achieve this, we incorporate a further-neighbor antiferromagnetic interaction, which we hypothesize plays a critical role in stabilizing the UVC state over other possible spin configurations, such as the staggered vector chirality (SVC) order. Furthermore, we will highlight several key features in physical quantities that can serve as experimental signatures to test the adequacy of our model and refine our understanding of the microscopic interactions driving the UVC order. 

\subsection{Additinal Heisenberg Coupling $J_x$}
In real physical systems, further-neighbor couplings are invariably present due to the extended nature of electronic wavefunctions. Even though they are typically significantly weaker than nearest-neighbor interactions, these additional interactions can significantly influence the magnetic ground state, subtly shifting the balance between competing orders and stabilizing exotic configurations. Here, we will incorporate further neighbor couplings to establish the UVC order in the MLL Heisenberg antiferromagnet. 

\begin{figure}[t]
\includegraphics[width=0.8\columnwidth]{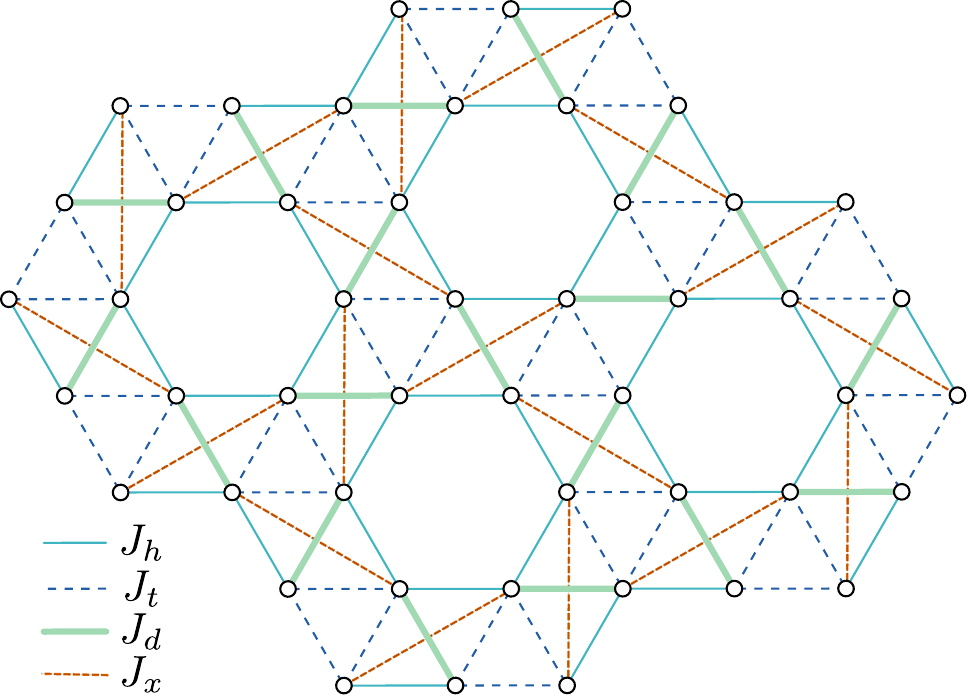}
\caption{The Maple-leaf lattice augmented by an additional further neighbor interaction, $J_x$, which engenders the UVC order ground state. This modified lattice is topologically equivalent to the ruby lattice.} \label{fig-latticeJx}
\end{figure}

Referring to Fig.~\ref{fig-lattice} (b), one can realize that adding a further neighbor antiferromagnetic coupling,  denoted as $J_x$, in this manner:
\begin{center}
  \includegraphics[width=0.4\columnwidth]{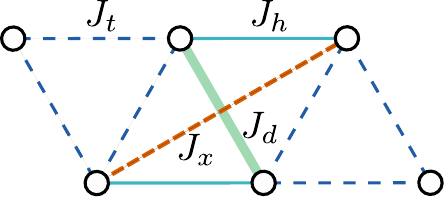}
\end{center}
does not alter the existing UVC state, but rather enhances its stability, as the spins astride these $J_x$ bonds are antiparallel. The inclusion of these bonds results in a lattice that becomes topologically equivalent to the ruby lattice, as illustrated in Fig.~\ref{fig-latticeJx}. Also, take note that the model exhibits self-duality under the exchange of $J_d$ and $J_x$, meaning the system remains invariant when these two coupling constants are swapped. The energy per spin of the UVC state in this modified model is given by
\be
e_{\text{UVC}} = -\frac{J_t}{2} + J_h \cos\frac{\pi}{3} - \frac{J_d}{2} - \frac{J_x}{2}.
\ee

\begin{figure}[t]
\includegraphics[width=0.9\columnwidth]{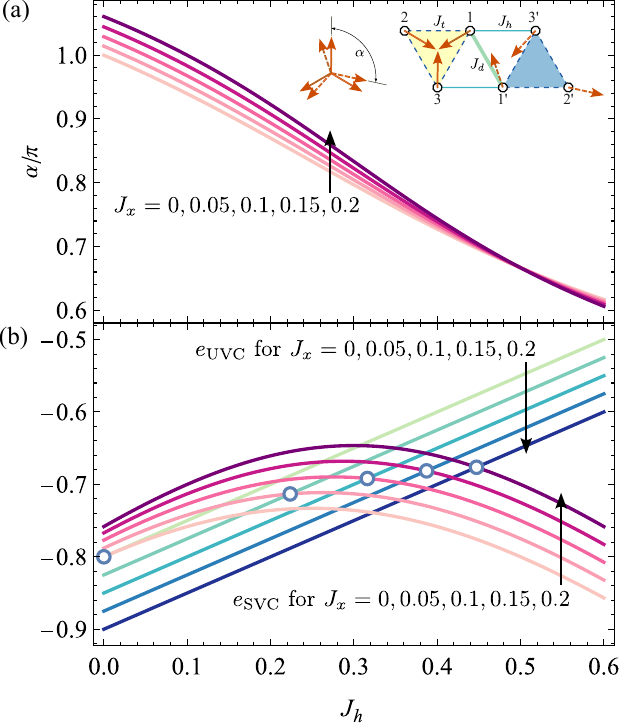}
\caption{(a) Variation of the canting angle, $\alpha$, for the SVC order as described by \eqref{eq:canting_2} for parameters $(J_d,J_t)=(1.0,0.6)$. (b) Comparison of the energy per spin for the SVC and UVC order in the $J_d$-$J_t$-$J_h$-$J_x$ model with $(J_d,J_t)=(1.0,0.6)$. The UVC is lowest in energy when $J_x/J_d>\left(J_h/J_d\right)^2$. Energy crossovers are indicated by open markers.}\label{fig-energy}
\end{figure}

For the SVC order, on the other hand, the energy evaluation is more involved, as the interactions $J_d$, $J_h$, and $J_x$ all compete to determine the canting angle, $\alpha$. To determine the $\alpha$ and to obtain the energy of the SVC order, we treat $\alpha$ as a variational parameter and minimize the energy with respect to it to derive the new expression for $\alpha$, which is now given by
\be\label{eq:canting_2}
\alpha = \pi + \arctan \left[\frac{\sqrt{3} (2J_h-J_x)}{2J_h-2J_d+J_x}\right]
\ee
[also see Fig.~\ref{fig-energy} (a)]. The energy per site for this SVC state, incorporating all four interactions, is then expressed as
\bea
e_{\text{SVC}} = -\frac{J_t}{2} &+& J_h \cos(\frac{2 \pi}{3} + \alpha) + \frac{J_d}{2} \cos\alpha\nonumber\\
&+&\frac{J_x}{2} \cos\left(\frac{2 \pi}{3} - \alpha\right).
\eea

The energies for the SVC and UVC orders can be compared [Fig.~\ref{fig-energy} (b)], revealing that the UVC becomes stricktly lower in energy when $$\frac{J_x}{J_d}>\left(\frac{J_h}{J_d}\right)^2.$$ This condition indicates that even a relatively weak $J_x$ is sufficient to engender the UVC order. To ensure the robustness of our findings and to confirm that no intermediate states exist between the SVC and UVC orders, we have performed iterative energy minimizations. These simulations confirm that the transition between SVC and UVC is direct and that no metastable or intermediate spin states intervene. 

\begin{figure*}[t]
\includegraphics[width=0.95\textwidth]{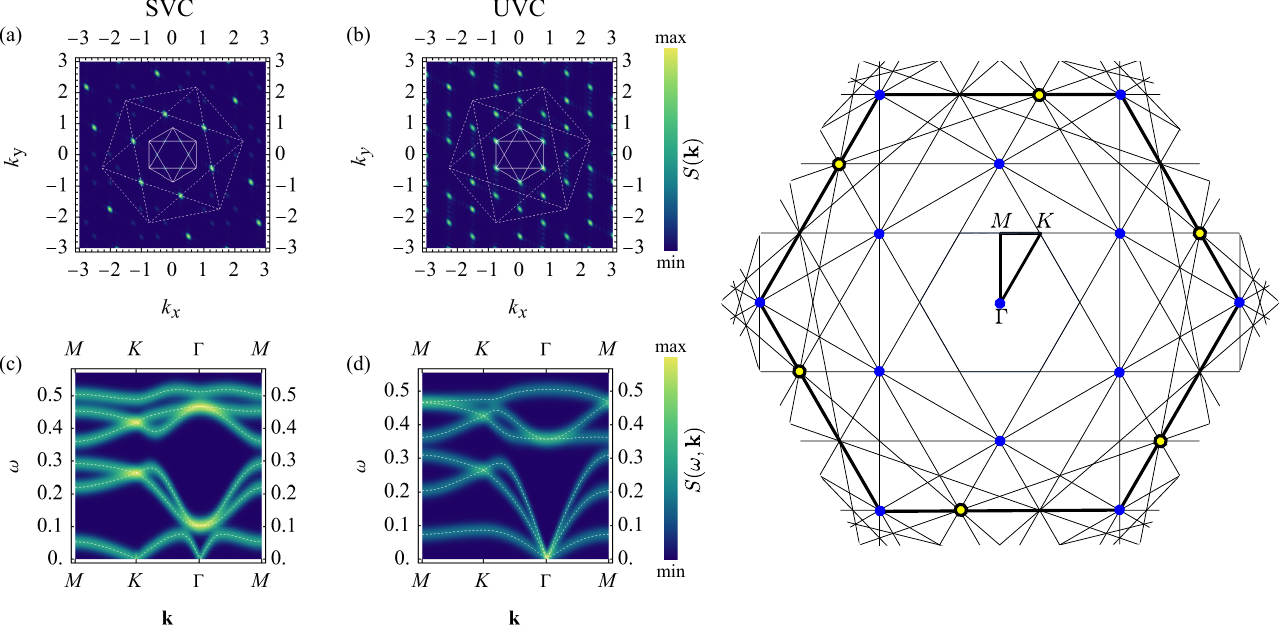}
\caption{The static spin structure factor for: (a) canted-$120^\circ$ or staggered vector chirality (SVC) order [$(J_d,J_t,J_h,J_x)=(1.0,0.4,0.2,0.03)$] (b) uniform vector chirality (UVC) order [$(J_d,J_t,J_h,J_x)=(1.0,0.4,0.2,0.05)$] on the Heisenberg model defined on the lattice depicted in Fig~\ref{fig-latticeJx}. The first three Brillouin zones of the MLL are indicated by solid white lines. The first three Brillouin zones of the corresponding triangular lattice, which is 1/7-th site depleted to reach the maple-leaf lattice, are marked with dashed lines. The dynamical spin structure factor for: (c) canted-$120^\circ$ or SVC order (d) UVC order, calculated for the same parameter values as in (a) and (b), respectively. The LSWT magnon branches are overlaid and depicted as white dashed lines. The right panel displays the further Brillouin zones of the MLL. The blue points denote the reciprocal lattice, where the yellow disks depict the positions of the Bragg peaks for the SVC state shown in (a).} \label{fig-Sq_UVC}
\end{figure*}

\subsection{Spin Structure Factor}
In the following, we calculate the static and dynamic spin structure factors for both the SVC and UVC order, which can be observed in neutron scattering experiments to ascertain the existence of the additional $J_x$ interactions. In the top panels panel of Fig.~\ref{fig-Sq_UVC}, we present the static spin structure factors,
$$
S(\mathbf{k})=\frac{1}{N}\sum_{i,j=1}^N\vec{S}_i\cdot\vec{S}_i\exp(-i\mathbf{k}.\vec{r}_{ij}),
$$
where $i,j$ runs over all $N$ spins, and $\vec{r}_{ij}$ denotes the vector connecting site $i$ and $j$, for both magnetic orders. The spin-structure factors for the two orders are markedly distinct. For the staggered vector chirality (SVC) order, the spin structure factor exhibits no Bragg peaks at the reciprocal lattice points, defined as $\mathbf{G} = n_1 \mathbf{b}_1 + n_2 \mathbf{b}_2$, where $\mathbf{b}_1$ and $\mathbf{b}_2$ are the primitive reciprocal lattice vectors corresponding to the lattice vectors, and $n_1$ and $n_2$ are integers. Instead, the peaks lie on the boundary of the ninth Brillouin zone at the positions indicated by the yellow points in the right panel of Fig.~\ref{fig-Sq_UVC}. In contrast, the UVC order displays Bragg peaks at the $M$ points of the extended Brillouin zone. Notably, the magnetic unit cell for the UVC state is the same as the crystal unit cell [Fig.~\ref{fig-lattice} (b)] resulting in magnetic peaks appearing at the same positions as structural peaks. Consequently, this order is characterized by the ordering wavevector $\mathbf{q}=(0,0)$ for which we obtain peaks at all points in the reciprocal space except when $n_1+n_2$ mod $5$ is zero. The same has been observed experimentally and serves compelling evidence supporting the UVC order. 

We now employ Linear Spin Wave Theory (LSWT) to investigate the dynamics of our classical Heisenberg spin system. LSTW provides an effective framework to study quantum and thermal fluctuations around an ordered magnetic ground state, particularly at low temperatures. In this setting, the spins are treated as continuous vectors of fixed length, and the spin wave approximation captures small deviations from the ground state configuration. The key assumption is that, at low but finite temperatures, thermal fluctuations are dominated by long-wavelength, low-energy excitations—spin waves. LSWT enables us to obtain the spin-wave spectrum, offering a bridge to experimental observables such as inelastic neutron scattering.

In our case, the LSWT is carried out following standard procedures with six different types of magnons~\cite{PhysRevB.65.224405}. We choose the quantization axis to align with the local orientation of the spins in the classical ground state. After applying the linear Holstein-Primakoff transformation, the scalar product $\vec{S}_i\cdot\vec{S}_j$ is replaced by a bosonic quadratic form:
\bea
\vec{S}_i\cdot\vec{S}_j \rightarrow\ & & S^2 \cos \Theta_{ij}
- S \cos\Theta_{ij}  \left( a_{i}^\dagger a_{i} + a_{j}^\dagger a_{j} \right) \nonumber \\
&+& \frac{S}{2} \left( \cos\ \Theta_{ij}  - 1 \right) \left( a_{i}^\dagger a_{j}^\dagger + a_{i} a_{j} \right) \nonumber \\
&+& \frac{S}{2} \left( \cos \Theta_{ij} + 1 \right) \left( a_{i}^\dagger a_{j} + a_{i} a_{j}^\dagger \right).
\eea
Here, $\Theta_{ij}$ represents the angle between the classical spin vectors at sites $i$ and $j$. The resulting spin-wave Hamiltonians are solved in the momentum space by performing a Bogoliubov transformation, yielding six spin-wave branches. 

For the SVC order, the spectrum consists of five optical branches and one acoustic branch, the latter touches zero at the $\Gamma$ and $K$ with distinct velocities. For the representative SVC order case, we use the coupling parameters $(J_d,J_t,J_h,J_x)=(1.0,0.4,0.2,0.03)$, with the corresponding LSWT spectrum shown in Fig.\ref{fig-Sq_UVC} (c). This behavior of the acoustic branch is a generic feature of the SVC order~\cite{PhysRevB.65.224405}. In contrast, the UVC order exhibits three optical and three acoustic branches, as depicted in Fig.\ref{fig-Sq_UVC} (d). For demonstations, we have chosen $(J_d,J_t,J_h,J_x)=(1.0,0.4,0.2,0.05)$ for the UVC order. The acoustic branches all touch zero linearly at the zone center, consistent with the ordering wavevector, $\mathbf{q}=(0,0)$. Generally, a gap exists between the acoustic and optical branches, as well as between the highest-energy optical branch and the other two optical branches.

We also compute the dynamical spin structure factor (transverse component)
$$
S(\mathbf{k},\omega)= \frac{1}{N}\sum_{i,j=1}^N\exp(-i\mathbf{k}.\vec{r}_{ij})\int_{-\infty}^\infty dte^{i\omega t} S^x_i (0) S^x_j(t)
$$
within the LSWT, which is directly linked to measurable quantities in neutron scattering experiments. To account for thermal effects, we perform this calculation at a finite temperature, incorporating a Bose-Einstein distribution factor to model the thermal population of magnon modes. Additionally, we replace the Dirac delta function with a Gaussian to represent finite resolution or lifetime effects. The differences in the dynamical spin structure factor between the staggered vector chirality (SVC) and uniform vector chirality (UVC) orders are evident in Fig.~\ref{fig-Sq_UVC}(c) and (d), providing a robust means to validate our model.
 
\subsection{Themal Properties}
To elucidate the behavior of the SVC and UVC orders at finite temperatures, we analyze their thermal properties using both LSWT and classical Monte Carlo (CMC) simulations to probe the specific heat, which serves as a key indicator of magnetic correlations, phase transitions, and low-energy excitations. 

\begin{figure}[t]
\includegraphics[width=\columnwidth]{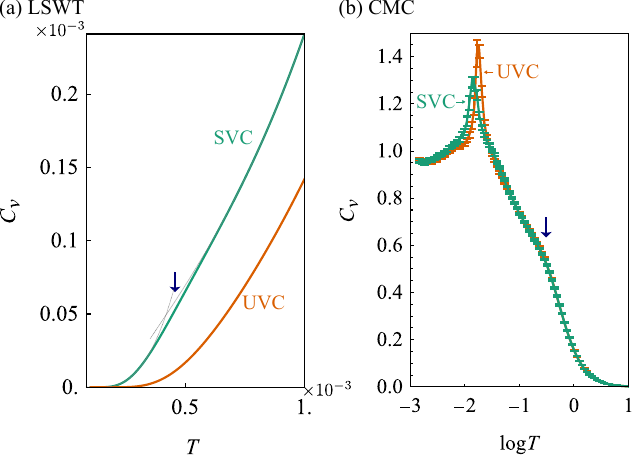}
\caption{The specific heat for the model calculated from (a) linear spin wave theory and (b) classical Monte Carlo (CMC) for both staggered vector chirality (SVC) and uniform vector chirality (UVC) orders. The interaction parameters are identical to those in Fig.~\ref{fig-Sq_UVC}. The Monte Carlo resulus shows an ensable average over 36 independent runs. The calculations for both orders show a sharp peak at low temperatures and a crossover at a higher temperature (marked by the arrow). The spin wave calculations finds that the SVC order shows an anomaly at a low temperature (marked by the arrow), whereas the UVC remains featureless at lower temperatures. The anomaly is caused by the small gap between the acoustic branch and the optical branches.}\label{fig-Cv_UVC}
\end{figure}

In Fig.~\ref{fig-Cv_UVC}, we present the specific heat of our model, comparing results from LSWT and CMC simulations. Panel (a) shows the specific heats of the SVC and the UVC order calculated from LSWT. The specific heat corresponding to the UVC order remains nearly featureless at low temperatures due to the presence of three strongly dispersing acoustic branches. In contrast, the SVC order exhibits an anomaly (indicated by the arrow) at a low temperature [see Fig. \ref{fig-Cv_UVC} (a)] resulting from the thermal activation of the lowest energy optical branch which is separated by a small gap from the acoustic branch [see Fig.\ref{fig-Sq_UVC} (c)].  

Panel (b) of Fig.~\ref{fig-Cv_UVC} shows results from the classical Monte Carlo simulations using simulated annealing for a fixed system size of $486$ spins ($9$ unit cells in each direction, commensurate with both orders) revealing two distinct features for both orders [see Fig.~\ref{fig-Cv_UVC} (b)]. First, a crossover at high temperatures (indicated by the arrow) possibly marking the development of short-range correlations. Second, a sharp peak at a lower temperature, $T_c \approx 0.02 \, \text{K}$, likely signals the onset of a short-range order. The UVC order displays a sharper peak compared to the SVC order.  This peak in specific heat is likely related to the stabilization of long-range magnetic order in \ce{Ho3ScO6}, which exhibits a $\lambda$-shaped peak in specific heat at $T_N \approx 4 \, \text{K}$, indicative of a second-order transition to an antiferromagnetic state. Since the two orders show no significant differences in the Monte Carlo simulations, we do not find it worthwhile to increase the system size further. However, we draw attention to the fact that, at the lowest temperature, both orders show a specific heat slightly below $1$, which may suggest the presence of soft modes\cite{PhysRevLett.68.855,Schnabel2012-ky,PhysRevB.78.094423}. This aspect will be investigated in more detail elsewhere. The observation of the sharp peak and the absence of anomalies at low temperatures suggests that our minimal model successfully captures the essential physics of \ce{Ho3ScO6}. Nevertheless, the actual phase transition observed in \ce{Ho3ScO6} points to the presence of weak interlayer couplings.

\section{Conclusions and Discussions}
In this article, we have investigated the classical antiferromagnetic Heisenberg model on the maple leaf lattice, in light of a newly reported relatization of the maple leaf magnetic system \ce{Ho3ScO6}~\cite{Lake2025}. The powder neutron scattering experiments suggest the appearance of a uniform vector chirality (UVC) magnetic order at low temperatures, which cannot be explained by a purely nearest-neighbor model. The maple leaf lattice has three symmetry inequivalent nearest neighbor models with couplings $J_d$, $J_t$, and $J_h$~\cite{Ghosh2022}. Previous theoretical studies have primarily focused on the case where $J_t = J_h$, where the system adopts a staggered vector chirality (SVC) magnetic order~\cite{PhysRevB.65.224405,Farnell2011,Farnell2014,Ghosh2022,Gresista2023,Ghosh2023}. Here, we extend beyond this specific condition and find that the SVC order remains stable across a broader region in the parameter space. However, we also find that the UVC order can not become the ground state within the nearest-neighbor model alone. We then introduce a plausible further-neighbor antiferromagnetic coupling, $J_x$, to find a weak $J_x$ is sufficient to compel the system to undergo a first-order phase transition from the SVC order and into the UVC order. Using linear spin wave theory, we calculate the dynamical spin structure factor for both magnetic orders, revealing starkly different behaviors that can be probed through neutron scattering experiments to validate our proposed model. Additionally, we compute the specific heat of the classical spin model, which exhibits close resemblance to the experimental observations in Ref.~\cite{Lake2025}.

The model presented in this article serves as a basic framework for understanding the maple leaf magnetic system \ce{Ho3ScO6}. We have not accounted for the effects of the crystal electric fields, which are expected to be prominent in the $4f$-electron systems. Given the centrosymmetric trigonal structure of \ce{Ho3ScO6}, nonzero coefficients for the Stevens Operators $O_2^0$, $O_4^0$, and $O_6^0$  are expected, which is likely to produce a planar anisotropy that further stabilizes the coplanar magnetic order~\cite{Koster1963-qd,Newman2009-wz}. Another interaction likely to be significant in \ce{Ho3ScO6} is the dipolar interaction, which is known to be crucial in rare-earth magnetic systems~\cite{Jensen1991,Maestro2004}. In the context of the maple-leaf lattice, the dipolar interaction warrants further theoretical investigation to fully understand its impact on the magnetic order.

It is also worth noting that the static structure factor alone cannot possibly distinguish whether the system adopts a UVC order or favors a particular state with uniform positive (or negative) vector chirality. With the inclusion of additional $J_x$ interactions, we have successfully established the UVC magnetic order. A typical way to promote a state with a preferred chirality is to include a Dzyaloshinskii–Moriya interaction (DMI). In our analysis, we introduced a DMI on the triangles that stabilizes the positive chirality order. We find that the primary effect of the DMI is to gap out one of the three acoustic modes, leaving only two Goldstone modes. However, we should highlight that for $4f$ ions in a centrosymmetric environment, the DMI is generally prohibited due to symmetry constraints. Nevertheless, we can not rule out the possibility that couplings to the conduction electrons may induce an effective, albeit weak, DMI~\cite{PhysRevB.110.054427,PhysRevB.96.115204}. In the absence of such an interaction, we can confidently assert that the maple leaf magnet \ce{Ho3ScO6} assumes a uniform vector chirality order, and the chiral symmetry is spontaneously broken at a finite temperature. 

Further investigations of \ce{Ho3ScO6}~\cite{Lake2025}, could focus on confirming the presence of the $J_x$ interactions via inelastic neutron scattering, the determination of crystal field levels, and conducting low-temperature measurements to detect whether the potential chirality-breaking transition conincides with the N\'eel transition seen at $T_N \approx 4 \, \text{K}$ or occurs at a lower temperature. Given that \ce{MgMn3O7.3H2O}, a spin-3/2 maple-leaf lattice antiferromagnet candidate, also exhibits magnetic ordering at low-temperatures~\cite{Haraguchi2018}, the insights presented here may also shed light into its behavior and this might also be an alluring candidate for further experimental and theoretical investigations.
\bibliography{Refs.bib}
\end{document}